\documentclass[a4paper, amsfonts, amssymb, amsmath, reprint, showkeys, aps, nofootinbib, twoside]{revtex4-1}
\usepackage[english]{babel}
\usepackage[utf8]{inputenc}
\usepackage{algorithm}
\usepackage{algpseudocode}
\usepackage{graphicx}
\usepackage{subfig}
\usepackage[colorinlistoftodos, color=green!40, prependcaption]{todonotes}
\usepackage{amsthm}
\usepackage{mathtools}
\usepackage{physics}
\usepackage{xcolor}
\usepackage{graphicx}
\usepackage[left=23mm,right=13mm,top=35mm,columnsep=15pt]{geometry} 
\usepackage{adjustbox}
\usepackage{placeins}
\usepackage[T1]{fontenc}
\usepackage{lipsum}
\usepackage{csquotes}
\usepackage{hyperref}
\usepackage[mathscr]{euscript}
\hypersetup{colorlinks=true, linkcolor=magenta, filecolor=cyan, citecolor=red,, urlcolor=blue}
\usepackage{array}
\newcommand{\PreserveBackslash}[1]{\let\temp=\\#1\let\\=\temp}
\newcolumntype{C}[1]{>{\PreserveBackslash\centering}p{#1}}
\newcolumntype{R}[1]{>{\PreserveBackslash\raggedleft}p{#1}}
\newcolumntype{L}[1]{>{\PreserveBackslash\raggedright}p{#1}}
\begin{document}
\title{Quantum-Inspired Optimization over Permutation Groups}

\author{Rathi Munukur}
    \affiliation{Research \& Advanced Engineering, Ford Motor Co., Dearborn, MI 48124, USA}
\author{Bhaskar Roy Bardhan}
    \affiliation{Research \& Advanced Engineering, Ford Motor Co., Dearborn, MI 48124, USA}
\author{Devesh Upadhyay}
    \affiliation{Research \& Advanced Engineering, Ford Motor Co., Dearborn, MI 48124, USA}  
\author{Joydip Ghosh}
    \email[Corresponding author. Email: ]{joydip.ghosh@gmail.com}
    \affiliation{Research \& Advanced Engineering, Ford Motor Co., Dearborn, MI 48124, USA}
    
\date{\today} 

\begin{abstract}
Quantum-inspired optimization (QIO) algorithms are computational techniques that emulate certain quantum mechanical effects on a classical hardware to tackle a class of optimization tasks. QIO methods have so far been employed to solve various binary optimization problems and a significant (polynomial) computational speedup over traditional techniques has also been reported. In this work, we develop an algorithmic framework, called \emph{Perm-QIO}, to tailor QIO tools to directly solve an arbitrary optimization problem, where the domain of the underlying cost function is defined over a permutation group. Such problems are not naturally recastable to a binary optimization and, therefore, are not necessarily within the scope of direct implementation of traditional QIO tools. We demonstrate the efficacy of Perm-QIO in leveraging the structure of cost-landscape to find high-quality solutions for a class of vehicle routing problems that belong to the category of non-trivial combinatorial optimization over the space of permutations.
\end{abstract}


\maketitle

\section{Introduction} \label{sec:introduction}
Quantum-inspired optimization (QIO) refers to a family of algorithms that classically emulates some quantum phenomena to tackle global optimization problems more efficiently than traditional optimization techniques~\cite{Swendsen1986PRL, Wolff1989PRL, Marinari1992EPL, Hukushima1996JPSJ, Katzgraber2006JSM, Machta2010PRE, Jarret2016PRA, Aramon2019FP,  Katzgraber2020OL}. The underlying objective function corresponding to an arbitrary non-trivial global optimization consists of many local minima often separated by tall and thin barriers. The performance of a global optimization solver, therefore, depends on how quickly it can hop out of such local optima to explore the global landscape. The time required to cross a barrier, however, scales exponentially with the barrier height for many traditional optimization methods, such as simulated annealing~\cite{Kirkpatrick1983science}. QIO proposes a solution to this challenge by emulating quantum effects, such as tunneling, which enables the QIO solver to sweep through the barriers as opposed to the computationally expensive thermal jumps exercised by simulated annealing. 

\begin{table}[htb]
\centering
    \begin{tabular}{ | C{2.3cm} || L{5.5cm} |}
    \hline
    & \\ 
    {\bf QIO tool} & {\bf Brief description} \\ \hline \hline
    & \\ 
    Parallel tempering (PT)~\cite{Swendsen1986PRL, Marinari1992EPL, Hukushima1996JPSJ} & Generalization of simulated annealing with multiple replicas maintained at different temperatures. Two neighboring (in the temperature space) replicas are exchanged periodically following a protocol. \\ 
     & \\ 
    \hline
     & \\ 
    Population annealing (PA)~\cite{Machta2010PRE} & Generalization of simulated annealing with multiple walkers maintained at a specific temperature. The population of walkers is resampled as temperature gets lowered eventually collapsing to a low-cost state. \\ 
     & \\ 
    \hline
     & \\ 
    Substochastic monte carlo (SSMC)~\cite{Jarret2016PRA} & Emulation of quantum annealing via the diffusion of population of walkers across the landscape of the objective function. Walkers get removed or created according to a protocol. \\
     & \\ 
    \hline
    \end{tabular}
\caption{\label{tab:qio}Some existing QIO tools developed to tackle discrete (specifically binary) optimization problems.}
\end{table}

There exists a number of techniques that can be characterized as members of QIO family. Table~\ref{tab:qio} shows a non-exhaustive list of existing QIO tools. These tools have so far been used to tackle binary optimization problems with an underlying objective function $f:\{0,1\}^n \rightarrow \mathbb{R}$, where $n$ is the number of binary variables. It is, in principle, possible to solve a broad class of discrete and combinatorial optimization problems using these QIO tools, once the problems are reformulated as unconstrained binary optimization tasks~\cite{Glover2019}. This reformulation process usually introduces significant overhead to the problem due to some additional constraints that are required to be satisfied. 

A class of combinatorial optimization problems that is of specific interest is optimization over a permutation group, where the domain of the underlying objective function is a set of permutations~\cite{BUCHHEIM2005DO, Ceberio2012}. A large number of computational problems, such as quadratic assignment problem~\cite{Koopmans1957}, linear ordering problem~\cite{Festa2001}, job-shop scheduling problem~\cite{Applegate1991OJC} and traveling salesman problem (TSP)~\cite{Kirkpatrick1983science}, belong to this category. Due to the combinatorial nature of the decision variables, not only are these problems NP-hard in general, but also approximation methods do not even exist for many of them~\cite{Sahni1976}. Optimization problems over a space of permutations, therefore, rely on metaheuristic approaches, such as QIO and evolutionary algorithms, as potential solvers for practical purposes. It is important to note that, for the purpose of this work we distinguish optimization problems from search problems. While it is in principle possible to leverage search algorithms, such as Tabu or guided local search, to tackle optimization problems, we assume that there exists some structure in the landscape of the objective function of an arbitrary (non-convex) optimization task that global optimization algorithms can leverage in order to find a good minima more efficiently using less number of queries.

The traditional implementation of QIO, when employed to tackle a permutation-based optimization, first maps the space of permutations to a space of binary permutation matrices and then imposes additional constraints to ensure that any optimal solution is valid and the resulting binary matrix can be mapped back to a well-defined permutation~\cite{Salehi2022}. Not only does this process consume a substantial computational resource, but it often fails to guarantee the validity of the output as a legitimate permutation for heuristic solvers, due to its extreme sensitivity on the penalty coefficients for the imposed constraint terms. Moreover, there exists a class of optimization problems over a set of permutations, where the cost of a given permutation depends on the entire sequence of symbols for that permutation itself, and therefore cannot be expressed as a sum of the products of elements between cost and the permutation matrices. These problems are \emph{not naturally recastable} to a binary optimization and thus are beyond the scope of direct implementation of traditional QIO tools~\cite{recastability}. We consider one such problem as a testbed to demonstrate the efficacy of our approach.

In this work, we develop an algorithmic framework, referred to as \emph{Perm-QIO}, that can adapt the requirements of QIO for the space of permutations. It thus enables existing QIO tools (except for the ones based on cluster updates), such as parallel tempering (PT), population annealing (PA) and substochastic monte carlo (SSMC), to directly tackle any optimization over permutations without resorting to a reformulation. With no loss of generality, we first describe our approach in the context of traveling salesman problem (as was done in the celebrated work by Kirkpatrick et al.~\cite{Kirkpatrick1983science}). We finally consider a vehicle routing problem, where one needs to find the specific optimal sequence of $n$ drop-off locations for a vehicle filled with $n$ parcels (each having different weights), such that the total energy consumption of the vehicle is minimum. The energy usage of the vehicle between two consecutive stops depends also on the total weight of the parcels carried, which varies along the route as the parcels get dropped off. This problem offers a non-trivial generalization of TSP, which cannot be recast as a binary optimization and, therefore, is not naturally amenable to traditional implementation of QIO tools.  

The remainder of the paper is organized as follows. In Sec.~\ref{sec:perm-opt}, we formulate the generic problem of optimization over a space of permutations. In Sec.~\ref{sec:qio}, we review the existing QIO tools and discuss how a traditional approach utilizes these tools to tackle binary optimization tasks. In Sec.~\ref{sec:perm-qio}, we elaborate Perm-QIO and describe how to implement this approach in the context of TSP. In Sec.~\ref{sec:eor}, we employ Perm-QIO for a vehicle routing problem with respect to minimal energy usage. We conclude in Sec.~\ref{sec:conclusions} with some potential future directions.
\section{Optimization over permutations} \label{sec:perm-opt}
In general an optimization problem is defined as follows:
Given an objective function $f:X \rightarrow {\mathbb R}$, find $x_\text{min} \in X$, s.t., $f(x_\text{min}) \leq f(x)$ $\forall x \in X$. In case of an optimization over permutations, $X \subseteq S_n$ for some $n \in {\mathbb N}$, where $S_n$ is the set of all permutations of the ordered symbols $\{1,2, \dots, n\}$. Since $|S_{n}|=n!$, the optimization over permutations gets intractable for brute-force methods as the number of symbols grow polynomially.

$S_n$ forms a \emph{group} (called the symmetric group) under composition of two permutations, with the following (non-exhaustive) generating sets~\cite{ash2013basic}:
\begin{enumerate}
  \item $S_n$ can be generated by the transpositions $(1,2)$, $(1,3)$, $\dots$, $(1,n)$.
  \item $S_n$ can be generated by the transpositions of adjacent symbols $(1,2)$, $(2,3)$, $\dots$, $(n-1,n)$.
\end{enumerate}

It is, in principle, possible to cast a class of permutation-based optimization as a 0/1-integer programming in order to use traditional solvers~\cite{conforti2014integer}. The first step towards this approach is to represent an arbitrary permutation $\sigma \in S_n$ as a binary permutation matrix $\Pi^{\sigma}$, such that,
\[
\Pi^{\sigma}_{ij} = \begin{cases}
1, & \text{if}\;\; \sigma(i)=j, \; \text{where}\; i,j \in \{1,2, \dots, n\}\\
0, & \text{otherwise.}
\end{cases}
\]
The optimization problem over permutations can then be formulated as a 0/1-integer programming:
\begin{equation} \label{eq:constr}
\begin{split}
\underset{x_{ij} \in \{0,1\}}{\text{min}}&f([x_{ij}]), \\
\text{subject to:} &\\
&\sum_{i}x_{ij}=1\;\; \forall j \in \{1,2, \dots, n\}, \\
&\sum_{j}x_{ij}=1\;\; \forall i \in \{1,2, \dots, n\},
\end{split}
\end{equation}
where $f$ is the cost function for a given permutation and the constraints enforce that the matrix $[x_{ij}]$ is a valid permutation matrix. The constraints remain same for all permutation-based optimization problems, whereas the cost function $f$ varies. For example, $f([x_{ij}]) = \sum_{ijk}c_{ij}x_{k,i}x_{k+1,j}$ for TSP, where $c_{ij}$ is the cost associated with the edge connecting $i^\text{th}$ and $j^\text{th}$ vertices. For another class of problems, the costs $c_{ij}$ depend non-trivially on the matrix $[x_{ij}]$, only if $[x_{ij}]$ corresponds to a valid permutation and must be set to $\infty$ otherwise. Whereas this construction is algorithmic, it does not necessarily allow a closed-form expression for $f([x_{ij}])$, and thus introduces significant additional complexity for traditional binary optimizers.

In order to solve permutation-based optimization problems using unconstrained optimization methods, such as QIO, the constraints in formulation~(\ref{eq:constr}) need to be baked into the objective function as,
\begin{equation} \label{eq:unconstr}
\begin{split}
\underset{x_{ij} \in \{0,1\}}{\text{min}} J({\bf x}) = f([x_{ij}]) +
w_{1}\sum_{j}\left(\sum_{i}x_{ij}-1\right)^{2} \\
+w_{2}\sum_{i}\left(\sum_{j}x_{ij}-1\right)^{2},
\end{split}
\end{equation}
where $w_1$ and $w_2$ are penalty coefficients required to convert the constrained optimization~(\ref{eq:constr}) to an unconstrained one. While such a reduction may not exist for a class of permutation-based optimization as we discussed, even when it does, it introduces two additional complexities to this approach:
\begin{enumerate}
  \item No heuristic solvers guarantee that the obtained optimal binary matrix is also a valid permutation matrix, unless the solver finds the global minimum.
  \item For $n$ symbols, the size of the domain changes from $n!$ to $2^{n(n-1)}$, and the latter scales $\mathcal{O}(2^{n})$ times faster than the former for large $n$. So the reformulation indeed offers an exponentially larger space to search for the global optima.
\end{enumerate}
These observations motivate us to explore the possibility for adapting QIO techniques directly to the space of permutations.

In order to quantify the performance of a heuristic (or approximate) optimizer over a space of permutations, here we propose a hardware-agnostic dimensionless performance vector $[\mathscr{S}, \mathscr{E}]$, where $\mathscr{S}$ denotes the \emph{span} that is defined as the ratio of the number of unique evaluations of the cost function to the total number of points in the landscape and $\mathscr{E}$ denotes the average error tolerance over many instances. A good optimizer is then characterized by a small span $\mathscr{S}$ as well as a small error $\mathscr{E}$, which altogether signify the efficacy of the optimizer in yielding a high-quality solution (lower error with respect to the global minimum) by exploring a comparatively smaller fraction of the entire landscape.
\section{Quantum-inspired optimization} \label{sec:qio}
In this section, we review some QIO tools that have been successfully employed to tackle hard binary optimization problems. For the purpose of this work, we exclude the mathematical rigor and describe the tools from the algorithmic standpoint. An arbitrary $n$-dimensional binary optimization problem can be denoted as: ${\displaystyle\min_{\textbf{x}}}f(\textbf{x})$, where $f:\{0, 1\}^{n} \rightarrow {\mathbb R}$. For an arbitrary non-convex binary optimization, the landscape of the objective function $f$ is rough containing multiple local minima often separated by tall and thin barriers. This poses additional complexity for any traditional solver, such as simulated annealing, which requires a significant time (exponential with the height of the barrier) to hop out of a local minimum. QIO tools attempt to emulate (on a classical hardware) quantum jumps, as opposed to thermal jumps, that can offer significant computational advantage for a wide variety of non-convex binary optimization problems. 

\subsection{Replica Exchange Monte Carlo}

We first describe the replica exchange Monte Carlo (also referred to as parallel tempering) algorithm that has been employed to overcome the barriers on rough landscapes by simulating a number of replicas maintained at different temperatures, which are essentially some hyper-parameters in the context of an optimization problem~\cite{Swendsen1986PRL, Marinari1992EPL, Hukushima1996JPSJ}.

The Parallel Tempering (PT) algorithm can be thought of as a generalization of the simulated annealing, where $M$ non-interacting replicas (trial solutions) of the system are simulated simultaneously at temperatures $ T_{1}, T_{2}, \ldots T_{M}$. After performing a fixed number of Monte Carlo sweeps (Metropolis updates) two replicas at neighboring temperatures are swapped with each other. This swap between the replicas is subject to an acceptance probability given by,
\begin{equation} 
\label{eq:acceptance_probability}
p \left (E_{i}, T_{i} \rightarrow E_{i+1}, T_{i+1} \right ) = \min \{1, \exp(\Delta \beta \Delta E)\},
\end{equation}
where $\Delta \beta=1/T_{i+1}-1/T_{i}$ is the difference between the inverse temperatures of the replicas and $\Delta E = E_{i+1}-E_{i}$ is the difference in energy of the two replicas at those temperatures. This probability satisfies the detailed balance conditions so that the algorithm converges to the equilibrium Gibbs distribution for each temperature.

An outline of the PT algorithm, as applied to a system consisting of binary variables in an optimization problem, is given below:

\begin{itemize}
\item \textbf{Initialize}: Initialize the replicas $X_{1}, X_{2}, \ldots X_{m}$ of the system at temperatures $T_{1}, T_{2},\ldots T_{m}$, where temperatures are in descending order.
\item \textbf{Sweep}: On each replica $X_{j}$, perform $n$ flips of the variables randomly. Such flips are done by randomly selecting a site, proposing a flip for the binary variable therein, accepting (or rejecting) the proposal as per the Metropolis criterion and finally repeating it $n$ times.
\item \textbf{Swap}: Swap $m-1$ number of replicas that are adjacent to each other in the temperature space subject to the acceptance probability given by~(\ref{eq:acceptance_probability}).
\item \textbf{Repeat}: Repeat the steps 2 and 3 above until some termination criterion is met based on convergence or maximum number of iterations.
\end{itemize}

For a given replica at a specific temperature, a random walk in the temperature space is induced by the swap move. Optimal performance of the PT algorithm in this space depends on how the temperatures of the replicas are distributed. Two commonly used temperature profiles include geometric and inverse linear temperature profiles~\cite{PhysRevE.100.043311}. Under the geometric temperature profile, the $i^\text{th}$ temperature $T_i$ is given by,
\begin{equation}
T_{i} = T_{1} \left(T_{m}/T_{1}\right)^\frac{i-1}{m-1}.    
\end{equation}
Under the inverse-linear scheme, the $i^\text{th}$ inverse temperature
\begin{equation}
\beta_{i} = \beta_{m}+(\beta_{1}-\beta_{m})\frac{i-1}{m-1}.    
\end{equation}
Other approaches, such as energy-based methods and feedback-optimized techniques, have also been proposed to set the temperatures for parallel tempering that can boost the performance of PT for many instances~\cite{PhysRevE.100.043311}. Unless stated otherwise, we use geometric spacing as our temperature profile for implementing PT.

\subsection{Population Annealing}

Population annealing can be thought of as a combination of the simulated annealing with a differential reproduction of the replicas. In a way very similar to simulated annealing, the temperature of the system is systematically lowered through a sequence of temperatures. However, unlike simulated annealing, at each temperature a population of replicas is resampled in order for the replicas to stay close to the equilibrium ensemble Gibbs distribution at that temperature. After this resampling, Monte Carlo sweeps on each replica are performed. If the original higher temperature population starts in an equilibrium ensemble in the Gibbs distribution at the inverse temperature, then the resampling process ensures the final population is also an equilibrium ensemble at the lower temperature at the end of the simulation.

The steps involved in the population annealing are outlined below:

\begin{itemize}
\item \textbf{Initialize}: Start with a population of $R$ independent replicas of the system in the uniform distribution at inverse temperature $\beta_{0}$.
 \item \textbf{Resampling}: To create an approximately equilibrated sample at $\beta_{i}>\beta_{i-1}$, resample replicas with their relative Boltzmann weights
 
\begin{equation}
\tau_{i}(E_{j})=\exp \left [-(\beta{i}-\beta_{i-1}) E_{j} \right ]/Q_{j},
\end{equation}

where
\begin{equation}
Q_{i}=\frac{1}{R_{i-1}} \sum_{j=1}^{R_{i-1}} \exp \left [-(\beta_{i}-\beta_{i-1} ) E_{j} \right ].
\end{equation} 
Here $E_{j}$ denotes the energy of the $j$-th replica and $R_k$ denotes the population size for the $k^\text{th}$ iteration (corresponding to the temperature $\beta_k$).

\begin{itemize}
\item For each replica $j$ in the population at inverse temperature $\beta_{i-1}$, draw a random number $r$ uniformly in [0,1) .
\item In the new population, the number of copies in the $j$-th replica is then taken to be 
\begin{equation}
r_{i}^{j}=\begin{cases}
\lfloor \hat{\tau_{i}} (E_{j}) \rfloor 
& \textrm{if}  ~ r> \hat{\tau_{i}} (E_{j})-\lfloor \hat{\tau_{i}} (E_{j}) \rfloor \\
\lfloor \hat{\tau_{i}} (E_{j}) \rfloor +1 & \textrm{otherwise},
\end{cases}
\end{equation}
where $\lfloor\ldots\rfloor$ denotes the floor function and $\hat{\tau_{i}}(E_{j})=\left (R/R_{i} \right ) \tau_{i} (E_{j})$ is normalized to keep the population size close to $R$. The above choice of the number of copies $r_{i}^{j}$ ensures that the mean of the random, non-negative integer $r_{i}^{j}$ is $ \hat{\tau_{i}} (E_{j})$ with its variance minimized and proportional to $\sqrt{R}$.
\item Calculate the new population size
\begin{equation}
R_{i}=\sum_{j} r_{i}^{j}
\end{equation}
\end{itemize}
\item \textbf{Perform Metropolis updates:} Update each replica by $N_{s}$ rounds of MCMC sweeps (Metropolis updates) at inverse temperature $\beta_{i}$. 
\item \textbf{Evaluate energy cost:} Estimate energies as population averages
\begin{equation}
\bar{e}=\frac{1}{R_{i}} \sum_{j} E_{j}/N,
\end{equation}
where $N$ is the number of the binary variables.
\item \textbf{Repeat:} Continue until final temperature $\beta_{f}$ is reached.
\end{itemize}

Population annealing has shown to be an efficient algorithm for simulating spin glasses~\cite{PhysRevE.92.063307}. In general, it serves as a method suitable for simulating the equilibrium states of systems with rough energy landscapes, providing comparable performance as the parallel tempering~\cite{PhysRevE.92.013303}.

\subsection{Substochastic Monte Carlo}

Substochastic Monte Carlo (SSMC) is a diffusion Monte Carlo algorithm that is tailored to simulate stoquastic adiabatic processes. This algorithm is inspired by adiabatic quantum computation and can be used to simulate the diffusion of walkers in the search space. The SSMC algorithm can be thought as a combination of this diffusion and resampling at each time step where the walkers are either eliminated or duplicated depending on how these walkers perform according to the cost function of a given optimization problem. The final distribution of the walkers, after running the algorithm for a specific optimization problem, will be proportional to the final ground state. In this sense, SSMC algorithm can be considered as a method for solving optimization problems~\cite{Jarret2016PRA}. As a classical optimization algorithm, SSMC performs competitively with respect to other heuristic classical solvers for MAX-$k$-SAT problem at $k=2,3,4$.

In the following, we describe the use of SSMC method for finding the ground state of an arbitrary Hamiltonian defined on a set of binary variables. This approach can be naturally extended to solve an arbitrary binary optimization problem, where the Hamiltonian gets replaced by an cost function. Consider a family of stoquastic Hamiltonians $H(s)$ which is parameterized by a schedule $s$, representing an adiabatic process, where the parameter $s(t)$ is varied from $0$ to $1$. The corresponding imaginary-time dynamics represents a continuous-time diffusion process or random walk of a set of walkers, where walkers can die. The following equation captures this diffusion process
\begin{equation}
\frac{d}{dt} \psi=-H(s(t)) \psi.
\end{equation}
The discretized time-evolution in this case is a substochastic Markov chain and the corresponding time-evolution operator $e^{-H(s) \Delta t}$ is a substochastic matrix when we consider sufficiently small time steps $\Delta t$ with respect to the gap between the ground and the first excited states. In this substochastic process, the total probability decreases as some walkers die.

For a given cost function associated with an optimization problem, we write the Hamiltonian $H(s)$ as

\begin{equation}
H(s)=a(s) L+b(s)W,
\end{equation}
where the cost function is represented by the diagonal operator $W=\textrm{diag} (w_{1},w_{2},\ldots)$, $L$ is the graph Laplacian, and $a(s)$, $b(s)$ are two monotonically decreasing and increasing functions over $s\in[0,1]$, respectively, which determines how the Hamiltonian is being swept from $L$ to $W$. Initially, when $s$ is small, this Hamiltonian is dominated by $L$ and the walkers widely explore the search space. However, as $s$ becomes larger, the Hamiltonian $H(s)$ is dominated by the cost function when the walkers start to become more concentrated in the search space region where the minima of the problem are likely to be found.

We now illustrate how to compute the probabilities of each of the random transitions the walkers can go through for a given binary optimization problem. Let a graph $G=(V,E)$ represent this optimization problem where $V$ and $E$ denote the vertex and edge sets, respectively. The combinatorial Laplacian corresponding to this graph is denoted by $L=D-A$, where $D$ is the diagonal operator for the degrees o the vertices and $A$ is the adjacency matrix of the graph. If the given problem has $n$ binary variables, we can represent it on a $n$-dimensional hypercube with $2^{n}$ vertices, where the vertices represent the possible states of the variables and the edges are formed only if the connecting vertices differ by a single bit. 

A given walker in this random walk process can go through any of the random transitions on this hypercube: step, stay, die or spawn a new walker. The probability of dying or spawning a new walker is determined by the cost function while the probabilities for stepping or staying are provided by the underlying Markovian dynamics of the process. We show this by computing the probabilities of these transitions in the outline of the SSMC algorithms as illustrated below~\cite{Jarret2016PRA}:

\begin{itemize}
\item \textbf{Initialize} Initialize the walkers in the ground state of $H(s=0)$, which is typically the uniform distribution.
\item \textbf{Execute Markov chain:} As $s$ is increased from 0 to 1, execute the Markov chain $\prod_{j} 1-H \left (s( t_{j}) \right ) \Delta t_{j}$, where $\Delta t_{j}$ is the $j$-th time step size and $t_{j}=\sum_{k=1}^{j} \Delta t_{j}$.

\item \textbf{Perform random transitions:} At time $t$, compute the value $s=s(t)$ according to a schedule, and a walker on the $j$-th vertex of the hypercube will do one of the following:

\begin{enumerate}
\item Step to a neighboring vertex with probability $a(s) \Delta t d_{j}$, where $d_{j}$ is the degree of the $j$-th vertex.
\item Stay at the $j$-th vertex with probability $1-a(s) \Delta t d_{j}-|b(s) \Delta t \left (w_{j}-{\langle}W{\rangle} _{t} \right )|$, where ${\langle}W{\rangle}_{t}$ is the population mean computed with respect to the current population of the walkers at time $t$.
\item If $ b(s) \Delta t \left (w_{j}-{\langle}W{\rangle} _{t} \right )>0$, the walker dies with this probability.
\item If $b(s) \Delta t \left (w_{j}-{\langle}W{\rangle} _{t} \right )<0$, then then the walker spawns a new one at the $j$-th vertex with probability $b(s) \Delta t \left ({\langle}W{\rangle} _{t}-w_{j} \right)$.
\end{enumerate}
\item \textbf{Repeat}: Repeat the above steps until convergence in energy is achieved or we run out of the walkers.
\end{itemize}

It is important to suitably replenish the population after each time step otherwise the process could terminate after a short amount of time. It has been shown that by setting an adaptive energy threshold, it is possible to replenish the population effectively. If the energy of the walker is above the energy threshold, it dies and if its energy lies below the threshold it spawns a new worker at its site. This replenishment strategy has been captured in the above outline of the algorithm. In addition, it is important to ensure that during the simulation, the population size remains close to a nominal value and it can be accomplished by adjusting the energy threshold ${\langle}W{\rangle}_{t}$, specifically by selecting an energy offset $E$ and replacing ${\langle}W{\rangle}_{t}$ by ${\langle}W{\rangle}_{t}-E$.

The above choice of probabilities, in conjunction with the adaptive replenishment of the population of the walkers, ensures that the initial distribution remains close to the quasistationary distribution which is proportional to the ground state of the original Hamiltonian. The optimum solution is found when at least one of the walker is found at the minimum energy vertex of the hypercube. The SSMC algorithm thus provides the ground state, or the lowest energy state as the solution of the binary optimization problem.
\section{Perm-QIO} \label{sec:perm-qio}
We observe that in order to solve an arbitrary optimization problem under traditional implementation of QIO tools, the following criteria are necessary (also sufficient for tools that do not require cluster updates):
\begin{enumerate}
  \item \emph{Drawing an arbitrary number of replicas randomly from a uniform distribution}:
  
  - In the context of a discrete optimization, this criterion reduces to a random sampling (from uniform distribution) for each variable within its domain. In case the domains of all variables are same, such as when the variables are all integers within a fixed interval, one can satisfy this condition either by drawing a random number for each variable within its domain or by drawing a random instance from the domain of the objective function. As an example, let's consider an objective function $f(x_{1},x_{2},x_{3})$, where $2 \leq x_{k} \leq 8$ for all $k \in \{1,2,3\}$. We can satisfy this criterion by generating a replica, where each $x_k$ is chosen randomly within $[2,8]$, or by choosing a random number between $222$ and $888$ and then assigning the $k^\text{th}$ digit of the number to the variable $x_k, \;\forall k \in \{1,2,3\}$.    
  \item \emph{Performing local operations on a given replica to transform it to another replica that is adjacent to the original one}:
  
  - In the context of binary optimization, this criterion boils down to single variable updates -- e.g., a single spin-flip for a Hamiltonian of a spin chain. One can adapt this criterion for an arbitrary discrete optimization, where a single variable needs to be updated by one of its two adjacent values. In any case, satisfying this criterion essentially means that QIO requires a concept of \emph{neighborhood} that must be defined on the domain of the objective function for any replica. The elements of a specific neighborhood are nearest neighbors connected via local operations.
\end{enumerate}

\subsection{Framework}
The primary challenge one requires to tackle in order to extend QIO directly to permutation-based optimization problems is to systematically adapt these criteria for a space of permutations. Here we describe how Perm-QIO achieves this goal:

\begin{figure}[htb]
\includegraphics[width=\linewidth]{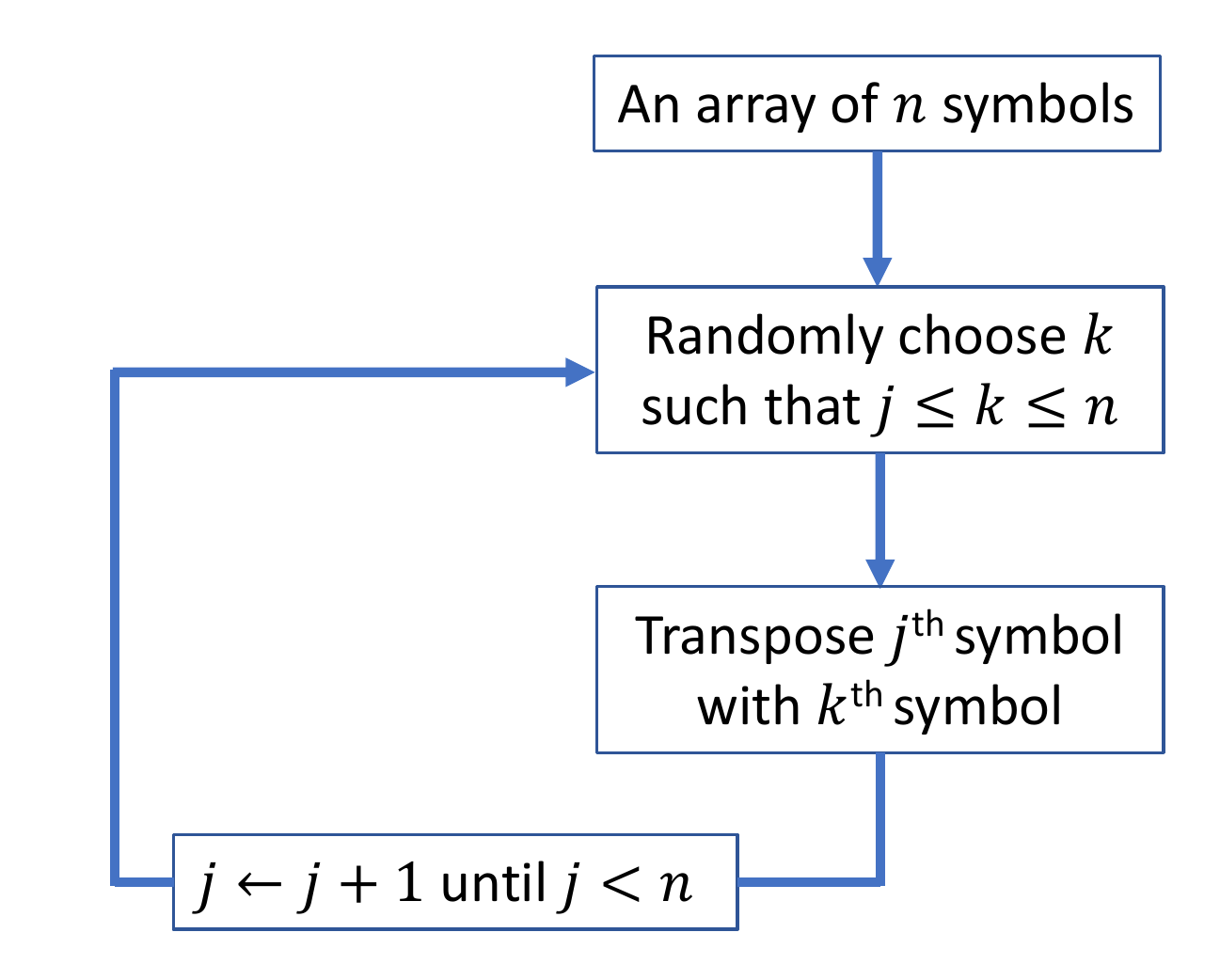}
\caption{Flowchart for Durstenfeld's implementation for Fisher-Yates algorithm to generate random permutations~\cite{Durstenfeld1964CA}.}
\label{fig:randomPerm}
\end{figure}

\begin{figure}[htb]
\includegraphics[width=\linewidth]{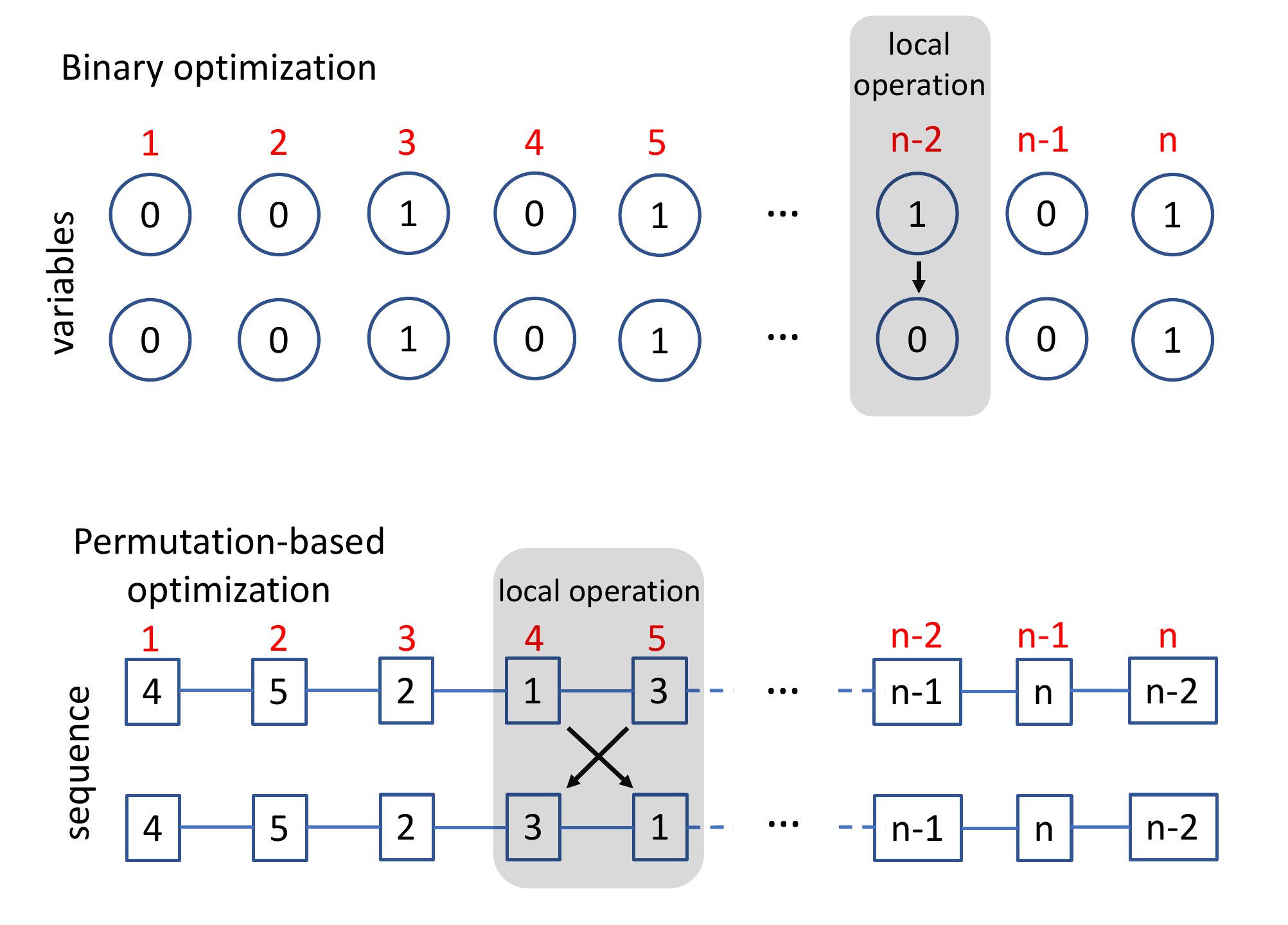}
\caption{Local operations for a generic binary optimization and optimization over permutation spaces. For a binary optimization it involves an update of a single variable selected randomly. For permutation-based optimization it involves an adjacent transposition for a consecutive pair selected randomly.}
\label{fig:adjTransposition}
\end{figure}

\subsubsection{Sampling from random permutations}
Generating a set of random replicas for a permutation-based optimization reduces to generating random permutations for the sequence of symbols, which was first proposed by Fisher \& Yates~\cite{FisherYates}. We use an efficient implementation of Fisher-Yates algorithm, as shown in Fig.~\ref{fig:randomPerm}, to generate the replicas sampled from a uniform distribution of permutations~\cite{Durstenfeld1964CA}. This step carries no additional computational overhead as the algorithm scales linearly with the number of symbols in the  sequence and can be parallelized easily on multiple cores to generate multiple replicas simultaneously. 

\subsubsection{Local operations over a space of permutations}
QIO requires defining a neighborhood for each replica and also a local operation that transforms one replica to another within its neighborhood. Whereas the local operations are supposed to perturb a replica minimally, it is also imperative that the entire domain of the underlying objective function must be generated from an arbitrary replica via the composition of such local operations. For a binary optimization problem, the neighborhood of a given replica is defined as the set of replicas obtained by flipping a single variable. Fig.~\ref{fig:adjTransposition} shows how a local operation acts on an arbitrary replica for a binary optimization problem (upper panel) and also how we adapt it to an optimization over permutation groups replacing the single-variable update with an adjacent transposition (lower panel). It is important to note that an adjacent transposition does qualify for a local operation, as it is not only the minimal change that can be imposed on a sequence of symbols, but also is a generator of the symmetry group that consists of all possible permutations. If we view the domain of the objective function as a graph with each replica being a vertex that is only connected to its neighboring replicas via edges, it is a hypercube for an arbitrary binary optimization problem, where the dimension of the hypercube is the same as the number of variables. For optimization over permutations of a (non-cyclic) sequence, the graph is still isomorphic to a hypercube with dimension $n-1$, where $n$ is the number of symbols in the sequence.

\subsection{Example: Traveling Salesman Problem}
The optimization version of the Traveling Salesman Problem (TSP) can be stated as follows: Given a set of $n$ points in a metric space (i.e., distance between every pair of points is known), find the shortest route that visits all the points exactly once and then returns to the starting point. The decision version of TSP is known to be NP-complete. In order to find the exact solution using brute-force, one needs to perform $n!$ operations to find the shortest route in the worst case. Dynamic programming offers a time-complexity of $\mathcal{O}(n^{2}2^{n})$ trading it off against a space-complexity of $\mathcal{O}(n2^{n})$. A number of approximation as well as heuristic algorithms, however, do exist to tackle TSP that can offer high-quality solutions more efficiently than exact algorithms~\cite{Rosenkrantz1977}.

\begin{figure}[htb]
\includegraphics[width=\linewidth]{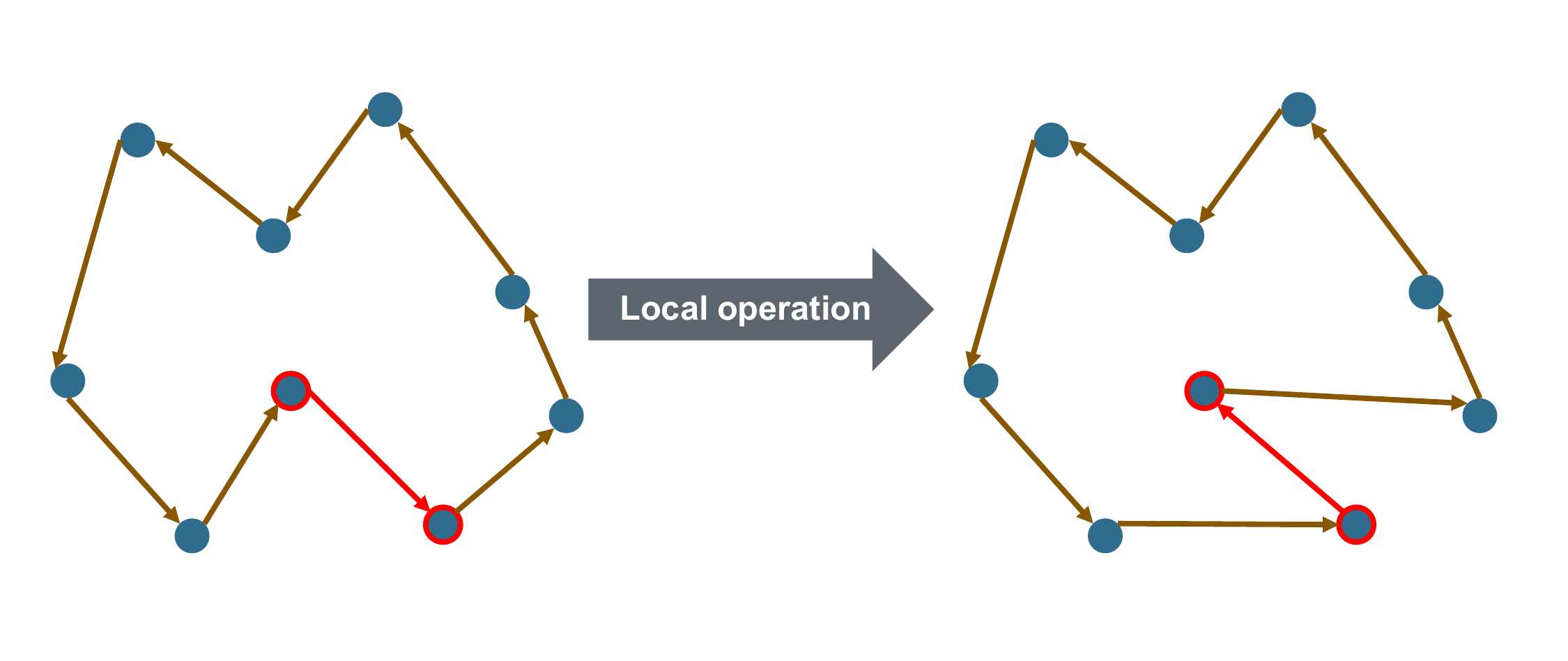}
\caption{A schematic diagram showing how an adjacent transposition transforms a given route for a TSP.}
\label{fig:TSP_AT}
\end{figure}

Here we use TSP as a candidate permutation-based optimization problem to describe Perm-QIO. The underlying objective function, under Perm-QIO formalism, is given by,
\begin{equation}
f_\text{TSP}(r) = \mathcal{D}[r],
\label{eq:f_tsp}
\end{equation}
where $r \in \mathcal{R}$ is a specific route and $\mathcal{D}:\mathcal{R} \rightarrow \mathbb{R}_{>0}$ denotes the mapping between the space of routes and the travel costs, such as the distance or time. With Perm-QIO, we can directly optimize $f_\text{TSP}$ in (\ref{eq:f_tsp}), as long as we have: 
\begin{enumerate}
    \item an efficient algorithm to sample randomly from $\mathcal{R}$ that can be achieved by generating random sequences of points using Durstenfeld's algorithm as outlined in Fig.~\ref{fig:randomPerm},
    \item a well-defined set of local operations for any $r \in \mathcal{R}$, which can be achieved by performing an adjacent transposition on two neighboring points for a given route, as shown in Fig.~\ref{eq:f_tsp}. Assuming the starting point to be the fixed one, the neighborhood of a given route $r$ with $n$ points (excluding the starting point) consists of $n-1$ possible routes corresponding to $n-1$ adjacent transpositions. 
\end{enumerate}

\section{Application to energy-optimal vehicle routing} \label{sec:eor}
In order to compare the performance of Perm-QIO over traditional implementation of QIO, we need a permutation-based optimization problem that can be served as a benchmark. What is a good benchmark for optimization over permutation groups? We argue that any permutation-based optimization that is hard (i.e., their decision version being NP-hard) with respect to exact, approximation and heuristic solutions can serve our purpose. TSP belongs to the class of permutation-based optimization problem and is NP-hard for exact solutions. However, there exist approximation and heuristic algorithms that can offer reasonably accurate solutions for TSP in polynomial time. Here, we formulate a constrained TSP, called Energy-optimal Single-vehicle Parcel Delivery Problem (ESPDP), for which the TSP heuristics cannot readily be employed to find the approximate solution. We use ESPDP to benchmark Perm-QIO against traditional QIO methods.

\subsection{ESPDP formulation}
ESPDP is an optimization problem, where one needs to find the optimal sequence of delivery stops for a vehicle under minimum energy consumption. The additional complexity with respect to TSP emerges from the fact that energy consumption here depends on the weight carried by the vehicle and as the vehicle drops the parcels its energy consumption gets lower. Therefore, unlike TSP, the cost to travel from the $i^{\text{th}}$ stop to the $j^{\text{th}}$ stop cannot be determined a priori for this case; it rather depends on the position of the route segment connecting the $i^{\text{th}}$ and the $j^{\text{th}}$ nodes in the entire route. Qualitatively speaking, the trade-off exists between dropping the heaviest parcel vs. serving the nearest or the farthest stop.

ESPDP satisfies the criteria for an optimization over a space of permutations, as the optimal solution exists in the set of permutation of stops. The objective function can be expressed as,
\begin{equation}
f_\text{ESPDP}(r) = \mathcal{C}[r],
\label{eq:f_espdp}
\end{equation}
where $r \in \mathcal{R}$ is a specific route and $\mathcal{C}:\mathcal{R} \rightarrow \mathbb{R}_{>0}$ denotes the map between the set of all possible routes and the total energy consumption. As discussed earlier, $\mathcal{C}[r]$ here cannot be expressed as $\sum_{i,j}C_{ij}(r_{ij})$, where $C_{ij}$ denotes some energy cost function for the route segment $r_{ij}$. $\mathcal{C}[r]$ must, therefore, be determined from a given route as,
\begin{equation}
\mathcal{C}[r] = \sum_{\substack{k=0 \\ {n+1}\coloneqq{0}}}^{n}\left[\left(W_\text{v}+\sum_{\substack{k < j \leq n}}W_{j}\right)C_{k,k+1}+\rho_{k,k+1}\right] 
\label{eq:C_r}
\end{equation}
where the route $r$ is defined via the sequence of stops $\{0, 1, 2, \dots, n, 0\}$, $W_{j}$ is the weight of the $j^\text{th}$ parcel (to be delivered at the $j^\text{th}$ stop), $W_\text{v}$ is the weight of the vehicle, $\rho_{i,j}$ captures the contributions from air resistance and $C_{i,j}$ is the proportionality coefficient connecting the weight of the vehicle to the fuel consumption cost for the route segment between $k^\text{th}$ stop and $(k+1)^\text{th}$ stop. We have considered this simplistic energy consumption model without any loss of generality, where the energy expenditure linearly depends on the weight of the vehicle. It is also important to note that, in absence of any fixed weight matrix, ESPDP (unlike TSP) cannot be readily recast to a binary optimization problem.

\subsection{Results: Performance of Perm-QIO for ESPDP}
\begin{figure}[htb]
\centering
\subfloat[]{{\includegraphics[width=\linewidth]{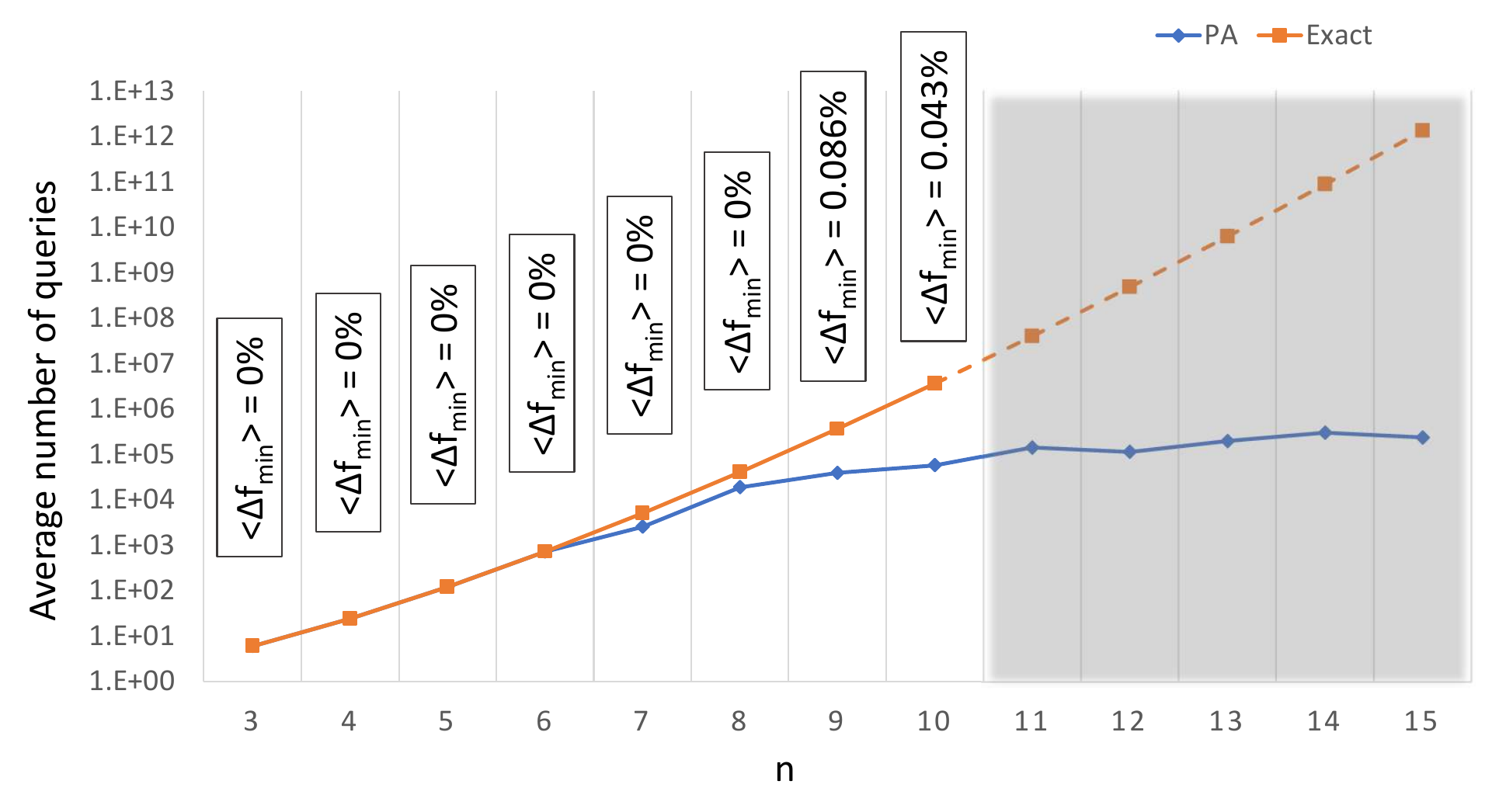} }\label{fig:result_PA}}

\subfloat[]{{\includegraphics[width=\linewidth]{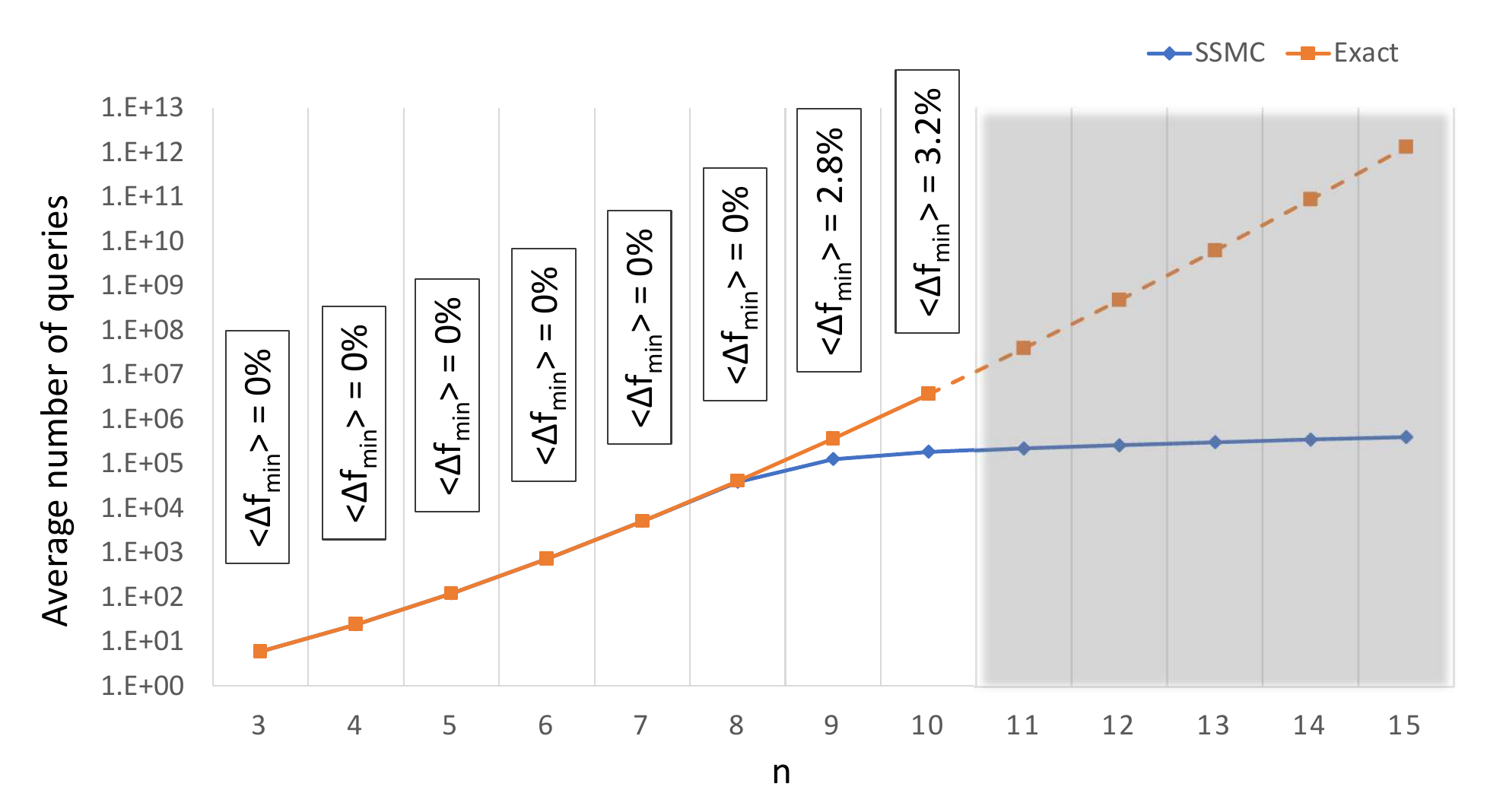} }\label{fig:result_SSMC}}

\subfloat[]{{\includegraphics[width=\linewidth]{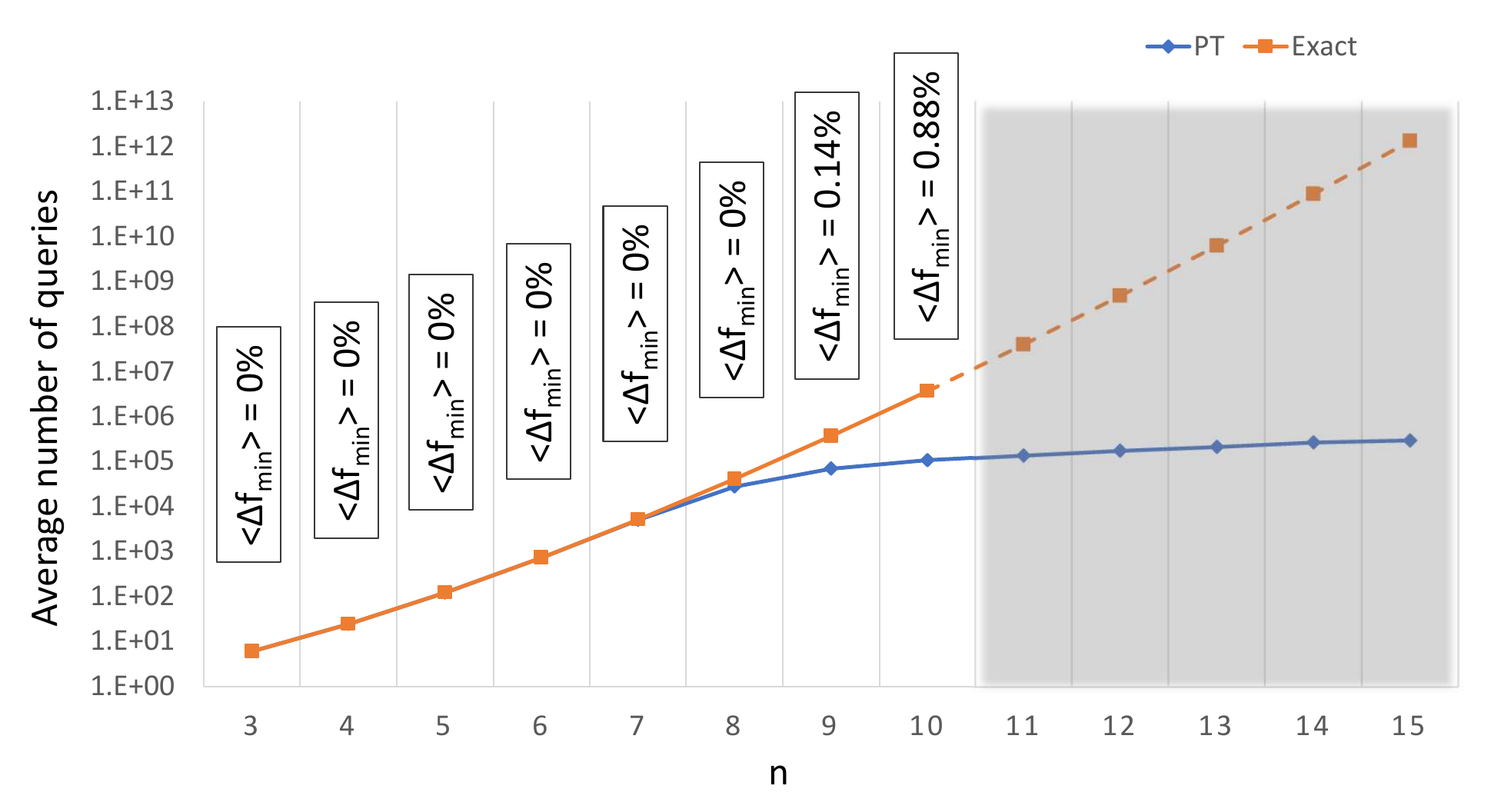} }\label{fig:result_PT}}

\caption{Average number of queries required to evaluate the objective function for an ESPDP having $n$ stops using (a) Perm-PA, (b) Perm-SSMC and (c) Perm-PT. $\langle{\Delta}f_\text{min}\rangle$ denotes the deviation from the global minima averaged over all the problem instances for each $n$. $\langle{\Delta}f_\text{min}\rangle$ has been computed for $n\leq10$, as the brute-force gets factorially more time-expensive as $n$ increases.}
\label{fig:results}
\end{figure}

In order to demonstrate the capabilities of Perm-QIO, we generate arbitrary instances of ESPDP for each $n$, where $n$ denotes the number of stops. Next we apply three different Perm-QIO tools: Perm-PA, Perm-SSMC and Perm-PT, which are essentially the adapted versions of PA, PT and SSMC algorithms for the permutation problems. We apply these tools on each problem instance for every $n$ and record the number of times the cost function is evaluated (or queried) by these algorithms. This metric can be considered as a hardware-agnostic measure of the temporal complexity. We also compute the deviation ${\Delta}f_\text{min}$ from the global minima for every instance, where the exact solution is obtained via the brute-force method for $n \leq 10$.

ESPDP is an NP-hard problem. Therefore, we are not expecting any exponential speedup here and the complexity of any heuristic algorithm must show the exponential scaling for a fixed amount of error tolerance, if $n$ is sufficiently large. The power of a suitable heuristic algorithm, however, is to defer this exponential behavior above a certain threshold, where the problem size is large enough. For our numerical experiments, we assume that we are below that threshold and set the hyperparameters of the Perm-QIO tools (once for all instances and problem sizes), such that the temporal scaling, measured by the average number of queries to the cost function, scales linearly with $n$.  

Among all the Perm-QIO tools, Perm-PA continues to run until a final temperature is reached and all the replicas converge to the same solution corresponding to that temperature. We, therefore, do not have a precise control over the number of queries for Perm-PA. However, the hyperparameters can be set such that the average number of queries scales approximately linearly with $n$, as shown in Fig.~\ref{fig:results}(a), where the vertical axis is on log-scale. We then set the hyperparameters of Perm-SSMC and Perm-PT, such that the average number of queries is comparable with that for the Perm-PA. For all $n \leq 10$, we compute the average deviation from the global minima, denoted by $\langle{\Delta}f_\text{min}\rangle$, where the global minima is calculated using the brute-force method that scales factorially with $n$. The results obtained from Perm-SSMC and Perm-PT are shown in Fig.~\ref{fig:results}(b) and \ref{fig:results}(c), respectively. The gray regions denote the regime for which we do not compute the global minima using the brute-force method due to its expensive runtime for each problem instance.

Our results demonstrate conspicuous efficacy of Perm-QIO tools in solving a hard optimization problem over the space of permutations that is not even naturally amenable to standard QIO. While Perm-SSMC and Perm-PT offer some room for improvement under appropriate hyperparameter tuning, the dimensionless performance vector (as defined in Sec.~\ref{sec:perm-opt}) obtained with Perm-PA for $n=10$ is $[0.016, 0.00043]$, which shows that on average it only explores $\sim1.6\%$ ($=57208/10!$) of the entire landscape of possible solutions for $n=10$ and yields an almost global minima, only differing from the true global minimum by $0.043\%$. This result also indicates that the underlying landscape of the cost function contains some structure that can be leveraged by an optimization heuristic in order to obtain the solution more efficiently compared to a regular search algorithm.
\section{Conclusions} \label{sec:conclusions}
 
 Quantum-inspired optimization has emerged as a powerful metaheuristic to tackle hard binary optimization problems. A certain class of combinatorial optimization problems, however, cannot naturally be recast to binary optimization. We have developed an algorithmic framework, referred to as Perm-QIO, that can leverage the logic behind traditional QIO tools to directly solve arbitrary combinatorial optimization problems, such as optimization over permutation groups. We also emphasize that many such problems can be viewed just as search problems, as opposed to optimization problems, and the global minima can be obtained through a search algorithm, such as Tabu or branch-and-bound. The importance of our work is that the underlying structure of the cost-landscape of an arbitrary optimization often makes the problems more suitable for an optimizer both in terms of efficiency and solution quality. A search algorithm, on the other hand, does not leverage the structure of the landscape to find the global (exact or approximate) minima and turns out to be less efficient for many problem instances with structured or semi-structured landscapes. 
 
 We consider a class of vehicle routing problems, called Energy-optimal single-vehicle parcel delivery problem or ESPDP, where a vehicle is scheduled to deliver packages on some drop-off locations under optimal usage of fuels. The non-trivial feature of this problem is the fact that the energy cost between two specific stops cannot be set a-priori, as the cost itself depends on the weight of the vehicle that changes with the parcels delivered en route. ESPDP is, therefore, a non-trivial combinatorial optimization problem over a space of permutations, where each different permutation (of stops) essentially describes a route. We argue that this problem is not naturally recastable to a binary optimization, and thus is not amenable to traditional QIO tools. This feature, in fact, makes ESPDP a suitable use case for Perm-QIO.
 
 We use ESPDP as a testbed to quantify the performance of Perm-QIO and benchmark against the exact brute-force method. Our results showed that Perm-QIO is capable of leveraging the structure of the cost-landscape of ESPDP and yield high-quality approximate solutions for most problem instances. The results obtained here encourage us to explore the capabilities of Perm-QIO for other hard combinatorial optimization problems and also to develop formal hyperparameter tuning methods that can ensure an optimal performance for such tools. These topics may be considered for future research.
\section*{Acknowledgements} \label{sec:acknowledgements}
The authors would like to thank Marwa Farag, Will Mallard and Owen Styles for useful discussions.
\newline

Data supporting the findings of this study are available on request.


\bibliography{QIO_Perm_Bib}

\begin{thebibliography}{27}%
\makeatletter
\providecommand \@ifxundefined [1]{%
 \@ifx{#1\undefined}
}%
\providecommand \@ifnum [1]{%
 \ifnum #1\expandafter \@firstoftwo
 \else \expandafter \@secondoftwo
 \fi
}%
\providecommand \@ifx [1]{%
 \ifx #1\expandafter \@firstoftwo
 \else \expandafter \@secondoftwo
 \fi
}%
\providecommand \natexlab [1]{#1}%
\providecommand \enquote  [1]{``#1''}%
\providecommand \bibnamefont  [1]{#1}%
\providecommand \bibfnamefont [1]{#1}%
\providecommand \citenamefont [1]{#1}%
\providecommand \href@noop [0]{\@secondoftwo}%
\providecommand \href [0]{\begingroup \@sanitize@url \@href}%
\providecommand \@href[1]{\@@startlink{#1}\@@href}%
\providecommand \@@href[1]{\endgroup#1\@@endlink}%
\providecommand \@sanitize@url [0]{\catcode `\\12\catcode `\$12\catcode
  `\&12\catcode `\#12\catcode `\^12\catcode `\_12\catcode `\%12\relax}%
\providecommand \@@startlink[1]{}%
\providecommand \@@endlink[0]{}%
\providecommand \url  [0]{\begingroup\@sanitize@url \@url }%
\providecommand \@url [1]{\endgroup\@href {#1}{\urlprefix }}%
\providecommand \urlprefix  [0]{URL }%
\providecommand \Eprint [0]{\href }%
\providecommand \doibase [0]{http://dx.doi.org/}%
\providecommand \selectlanguage [0]{\@gobble}%
\providecommand \bibinfo  [0]{\@secondoftwo}%
\providecommand \bibfield  [0]{\@secondoftwo}%
\providecommand \translation [1]{[#1]}%
\providecommand \BibitemOpen [0]{}%
\providecommand \bibitemStop [0]{}%
\providecommand \bibitemNoStop [0]{.\EOS\space}%
\providecommand \EOS [0]{\spacefactor3000\relax}%
\providecommand \BibitemShut  [1]{\csname bibitem#1\endcsname}%
\let\auto@bib@innerbib\@empty
\bibitem [{\citenamefont {Swendsen}\ and\ \citenamefont
  {Wang}(1986)}]{Swendsen1986PRL}%
  \BibitemOpen
  \bibfield  {author} {\bibinfo {author} {\bibfnamefont {R.~H.}\ \bibnamefont
  {Swendsen}}\ and\ \bibinfo {author} {\bibfnamefont {J.-S.}\ \bibnamefont
  {Wang}},\ }\href {\doibase 10.1103/PhysRevLett.57.2607} {\bibfield  {journal}
  {\bibinfo  {journal} {Phys. Rev. Lett.}\ }\textbf {\bibinfo {volume} {57}},\
  \bibinfo {pages} {2607} (\bibinfo {year} {1986})}\BibitemShut {NoStop}%
\bibitem [{\citenamefont {Wolff}(1989)}]{Wolff1989PRL}%
  \BibitemOpen
  \bibfield  {author} {\bibinfo {author} {\bibfnamefont {U.}~\bibnamefont
  {Wolff}},\ }\href {\doibase 10.1103/PhysRevLett.62.361} {\bibfield  {journal}
  {\bibinfo  {journal} {Phys. Rev. Lett.}\ }\textbf {\bibinfo {volume} {62}},\
  \bibinfo {pages} {361} (\bibinfo {year} {1989})}\BibitemShut {NoStop}%
\bibitem [{\citenamefont {Marinari}\ and\ \citenamefont
  {Parisi}(1992)}]{Marinari1992EPL}%
  \BibitemOpen
  \bibfield  {author} {\bibinfo {author} {\bibfnamefont {E.}~\bibnamefont
  {Marinari}}\ and\ \bibinfo {author} {\bibfnamefont {G.}~\bibnamefont
  {Parisi}},\ }\href {\doibase 10.1209/0295-5075/19/6/002} {\bibfield
  {journal} {\bibinfo  {journal} {Europhysics Letters ({EPL})}\ }\textbf
  {\bibinfo {volume} {19}},\ \bibinfo {pages} {451} (\bibinfo {year}
  {1992})}\BibitemShut {NoStop}%
\bibitem [{\citenamefont {Hukushima}\ and\ \citenamefont
  {Nemoto}(1996)}]{Hukushima1996JPSJ}%
  \BibitemOpen
  \bibfield  {author} {\bibinfo {author} {\bibfnamefont {K.}~\bibnamefont
  {Hukushima}}\ and\ \bibinfo {author} {\bibfnamefont {K.}~\bibnamefont
  {Nemoto}},\ }\href {\doibase 10.1143/JPSJ.65.1604} {\bibfield  {journal}
  {\bibinfo  {journal} {Journal of the Physical Society of Japan}\ }\textbf
  {\bibinfo {volume} {65}},\ \bibinfo {pages} {1604} (\bibinfo {year}
  {1996})}\BibitemShut {NoStop}%
\bibitem [{\citenamefont {Katzgraber}\ \emph {et~al.}(2006)\citenamefont
  {Katzgraber}, \citenamefont {Trebst}, \citenamefont {Huse},\ and\
  \citenamefont {Troyer}}]{Katzgraber2006JSM}%
  \BibitemOpen
  \bibfield  {author} {\bibinfo {author} {\bibfnamefont {H.~G.}\ \bibnamefont
  {Katzgraber}}, \bibinfo {author} {\bibfnamefont {S.}~\bibnamefont {Trebst}},
  \bibinfo {author} {\bibfnamefont {D.~A.}\ \bibnamefont {Huse}}, \ and\
  \bibinfo {author} {\bibfnamefont {M.}~\bibnamefont {Troyer}},\ }\href
  {\doibase 10.1088/1742-5468/2006/03/p03018} {\bibfield  {journal} {\bibinfo
  {journal} {Journal of Statistical Mechanics: Theory and Experiment}\ }\textbf
  {\bibinfo {volume} {2006}},\ \bibinfo {pages} {P03018} (\bibinfo {year}
  {2006})}\BibitemShut {NoStop}%
\bibitem [{\citenamefont {Machta}(2010)}]{Machta2010PRE}%
  \BibitemOpen
  \bibfield  {author} {\bibinfo {author} {\bibfnamefont {J.}~\bibnamefont
  {Machta}},\ }\href {\doibase 10.1103/PhysRevE.82.026704} {\bibfield
  {journal} {\bibinfo  {journal} {Phys. Rev. E}\ }\textbf {\bibinfo {volume}
  {82}},\ \bibinfo {pages} {026704} (\bibinfo {year} {2010})}\BibitemShut
  {NoStop}%
\bibitem [{\citenamefont {Jarret}\ \emph {et~al.}(2016)\citenamefont {Jarret},
  \citenamefont {Jordan},\ and\ \citenamefont {Lackey}}]{Jarret2016PRA}%
  \BibitemOpen
  \bibfield  {author} {\bibinfo {author} {\bibfnamefont {M.}~\bibnamefont
  {Jarret}}, \bibinfo {author} {\bibfnamefont {S.~P.}\ \bibnamefont {Jordan}},
  \ and\ \bibinfo {author} {\bibfnamefont {B.}~\bibnamefont {Lackey}},\ }\href
  {\doibase 10.1103/PhysRevA.94.042318} {\bibfield  {journal} {\bibinfo
  {journal} {Phys. Rev. A}\ }\textbf {\bibinfo {volume} {94}},\ \bibinfo
  {pages} {042318} (\bibinfo {year} {2016})}\BibitemShut {NoStop}%
\bibitem [{\citenamefont {Aramon}\ \emph {et~al.}(2019)\citenamefont {Aramon},
  \citenamefont {Rosenberg}, \citenamefont {Valiante}, \citenamefont
  {Miyazawa}, \citenamefont {Tamura},\ and\ \citenamefont
  {Katzgraber}}]{Aramon2019FP}%
  \BibitemOpen
  \bibfield  {author} {\bibinfo {author} {\bibfnamefont {M.}~\bibnamefont
  {Aramon}}, \bibinfo {author} {\bibfnamefont {G.}~\bibnamefont {Rosenberg}},
  \bibinfo {author} {\bibfnamefont {E.}~\bibnamefont {Valiante}}, \bibinfo
  {author} {\bibfnamefont {T.}~\bibnamefont {Miyazawa}}, \bibinfo {author}
  {\bibfnamefont {H.}~\bibnamefont {Tamura}}, \ and\ \bibinfo {author}
  {\bibfnamefont {H.~G.}\ \bibnamefont {Katzgraber}},\ }\href {\doibase
  10.3389/fphy.2019.00048} {\bibfield  {journal} {\bibinfo  {journal}
  {Frontiers in Physics}\ }\textbf {\bibinfo {volume} {7}} (\bibinfo {year}
  {2019}),\ 10.3389/fphy.2019.00048}\BibitemShut {NoStop}%
\bibitem [{\citenamefont {Zheng~Zhu}\ and\ \citenamefont
  {Katzgraber}(2020)}]{Katzgraber2020OL}%
  \BibitemOpen
  \bibfield  {author} {\bibinfo {author} {\bibfnamefont {C.~F.}\ \bibnamefont
  {Zheng~Zhu}}\ and\ \bibinfo {author} {\bibfnamefont {H.~G.}\ \bibnamefont
  {Katzgraber}},\ }\href {\doibase 10.1007/s11590-020-01570-7} {\bibfield
  {journal} {\bibinfo  {journal} {Optimization Letters}\ }\textbf {\bibinfo
  {volume} {14}},\ \bibinfo {pages} {2495} (\bibinfo {year}
  {2020})}\BibitemShut {NoStop}%
\bibitem [{\citenamefont {Kirkpatrick}\ \emph {et~al.}(1983)\citenamefont
  {Kirkpatrick}, \citenamefont {Gelatt},\ and\ \citenamefont
  {Vecchi}}]{Kirkpatrick1983science}%
  \BibitemOpen
  \bibfield  {author} {\bibinfo {author} {\bibfnamefont {S.}~\bibnamefont
  {Kirkpatrick}}, \bibinfo {author} {\bibfnamefont {C.~D.}\ \bibnamefont
  {Gelatt}}, \ and\ \bibinfo {author} {\bibfnamefont {M.~P.}\ \bibnamefont
  {Vecchi}},\ }\href {\doibase 10.1126/science.220.4598.671} {\bibfield
  {journal} {\bibinfo  {journal} {Science}\ }\textbf {\bibinfo {volume}
  {220}},\ \bibinfo {pages} {671} (\bibinfo {year} {1983})}\BibitemShut
  {NoStop}%
\bibitem [{\citenamefont {Glover}\ \emph {et~al.}(2019)\citenamefont {Glover},
  \citenamefont {Kochenberger},\ and\ \citenamefont {Du}}]{Glover2019}%
  \BibitemOpen
  \bibfield  {author} {\bibinfo {author} {\bibfnamefont {F.}~\bibnamefont
  {Glover}}, \bibinfo {author} {\bibfnamefont {G.}~\bibnamefont
  {Kochenberger}}, \ and\ \bibinfo {author} {\bibfnamefont {Y.}~\bibnamefont
  {Du}},\ }\href {\doibase 10.1007/s10288-019-00424-y} {\bibfield  {journal}
  {\bibinfo  {journal} {4OR}\ }\textbf {\bibinfo {volume} {17}},\ \bibinfo
  {pages} {335} (\bibinfo {year} {2019})}\BibitemShut {NoStop}%
\bibitem [{\citenamefont {Buchheim}\ and\ \citenamefont
  {J\"{u}nger}(2005)}]{BUCHHEIM2005DO}%
  \BibitemOpen
  \bibfield  {author} {\bibinfo {author} {\bibfnamefont {C.}~\bibnamefont
  {Buchheim}}\ and\ \bibinfo {author} {\bibfnamefont {M.}~\bibnamefont
  {J\"{u}nger}},\ }\href {\doibase
  https://doi.org/10.1016/j.disopt.2005.08.005} {\bibfield  {journal} {\bibinfo
   {journal} {Discrete Optimization}\ }\textbf {\bibinfo {volume} {2}},\
  \bibinfo {pages} {308} (\bibinfo {year} {2005})}\BibitemShut {NoStop}%
\bibitem [{\citenamefont {Ceberio}\ \emph {et~al.}(2012)\citenamefont
  {Ceberio}, \citenamefont {Irurozki}, \citenamefont {Mendiburu},\ and\
  \citenamefont {Lozano}}]{Ceberio2012}%
  \BibitemOpen
  \bibfield  {author} {\bibinfo {author} {\bibfnamefont {J.}~\bibnamefont
  {Ceberio}}, \bibinfo {author} {\bibfnamefont {E.}~\bibnamefont {Irurozki}},
  \bibinfo {author} {\bibfnamefont {A.}~\bibnamefont {Mendiburu}}, \ and\
  \bibinfo {author} {\bibfnamefont {J.~A.}\ \bibnamefont {Lozano}},\ }\href
  {\doibase 10.1007/s13748-011-0005-3} {\bibfield  {journal} {\bibinfo
  {journal} {Progress in Artificial Intelligence}\ }\textbf {\bibinfo {volume}
  {1}},\ \bibinfo {pages} {103} (\bibinfo {year} {2012})}\BibitemShut {NoStop}%
\bibitem [{\citenamefont {Koopmans}\ and\ \citenamefont
  {Beckmann}(1957)}]{Koopmans1957}%
  \BibitemOpen
  \bibfield  {author} {\bibinfo {author} {\bibfnamefont {T.~C.}\ \bibnamefont
  {Koopmans}}\ and\ \bibinfo {author} {\bibfnamefont {M.}~\bibnamefont
  {Beckmann}},\ }\href {http://www.jstor.org/stable/1907742} {\bibfield
  {journal} {\bibinfo  {journal} {Econometrica}\ }\textbf {\bibinfo {volume}
  {25}},\ \bibinfo {pages} {53} (\bibinfo {year} {1957})}\BibitemShut {NoStop}%
\bibitem [{\citenamefont {Festa}(2001)}]{Festa2001}%
  \BibitemOpen
  \bibfield  {author} {\bibinfo {author} {\bibfnamefont {P.}~\bibnamefont
  {Festa}},\ }\enquote {\bibinfo {title} {Linear ordering problem},}\ in\ \href
  {\doibase 10.1007/0-306-48332-7_260} {\emph {\bibinfo {booktitle}
  {Encyclopedia of Optimization}}},\ \bibinfo {editor} {edited by\ \bibinfo
  {editor} {\bibfnamefont {C.~A.}\ \bibnamefont {Floudas}}\ and\ \bibinfo
  {editor} {\bibfnamefont {P.~M.}\ \bibnamefont {Pardalos}}}\ (\bibinfo
  {publisher} {Springer US},\ \bibinfo {address} {Boston, MA},\ \bibinfo {year}
  {2001})\ pp.\ \bibinfo {pages} {1274--1276}\BibitemShut {NoStop}%
\bibitem [{\citenamefont {Applegate}\ and\ \citenamefont
  {Cook}(1991)}]{Applegate1991OJC}%
  \BibitemOpen
  \bibfield  {author} {\bibinfo {author} {\bibfnamefont {D.}~\bibnamefont
  {Applegate}}\ and\ \bibinfo {author} {\bibfnamefont {W.}~\bibnamefont
  {Cook}},\ }\href {\doibase 10.1287/ijoc.3.2.149} {\bibfield  {journal}
  {\bibinfo  {journal} {ORSA Journal on Computing}\ }\textbf {\bibinfo {volume}
  {3}},\ \bibinfo {pages} {149} (\bibinfo {year} {1991})},\ \Eprint
  {http://arxiv.org/abs/https://doi.org/10.1287/ijoc.3.2.149}
  {https://doi.org/10.1287/ijoc.3.2.149} \BibitemShut {NoStop}%
\bibitem [{\citenamefont {Sahni}\ and\ \citenamefont
  {Gonzalez}(1976)}]{Sahni1976}%
  \BibitemOpen
  \bibfield  {author} {\bibinfo {author} {\bibfnamefont {S.}~\bibnamefont
  {Sahni}}\ and\ \bibinfo {author} {\bibfnamefont {T.}~\bibnamefont
  {Gonzalez}},\ }\href {\doibase 10.1145/321958.321975} {\bibfield  {journal}
  {\bibinfo  {journal} {J. ACM}\ }\textbf {\bibinfo {volume} {23}},\ \bibinfo
  {pages} {555–565} (\bibinfo {year} {1976})}\BibitemShut {NoStop}%
\bibitem [{\citenamefont {Salehi}\ \emph {et~al.}(2022)\citenamefont {Salehi},
  \citenamefont {Glos},\ and\ \citenamefont {Miszczak}}]{Salehi2022}%
  \BibitemOpen
  \bibfield  {author} {\bibinfo {author} {\bibfnamefont {{\"O}.}~\bibnamefont
  {Salehi}}, \bibinfo {author} {\bibfnamefont {A.}~\bibnamefont {Glos}}, \ and\
  \bibinfo {author} {\bibfnamefont {J.~A.}\ \bibnamefont {Miszczak}},\ }\href
  {\doibase 10.1007/s11128-021-03405-5} {\bibfield  {journal} {\bibinfo
  {journal} {Quantum Information Processing}\ }\textbf {\bibinfo {volume}
  {21}},\ \bibinfo {pages} {67} (\bibinfo {year} {2022})}\BibitemShut {NoStop}%
\bibitem [{rec()}]{recastability}%
  \BibitemOpen
  \href@noop {} {}\bibinfo {note} {Throughout this work, ``\emph{not naturally
  recastable to binary optimization}'' means that the cost function for such a
  problem cannot be written as a closed-form (such as quadratic or even
  polynomial unconstrained binary optimization) analytic expression over the
  set of binary decision variables. The cost functions nevertheless can be
  algorithmically constructed for any arbitrary binary input permutation
  matrix, which, however, introduces additional redundancies to such
  problems.}\BibitemShut {Stop}%
\bibitem [{\citenamefont {Ash}(2013)}]{ash2013basic}%
  \BibitemOpen
  \bibfield  {author} {\bibinfo {author} {\bibfnamefont {R.}~\bibnamefont
  {Ash}},\ }\href {https://books.google.com/books?id=5m7CAgAAQBAJ} {\emph
  {\bibinfo {title} {Basic Abstract Algebra: For Graduate Students and Advanced
  Undergraduates}}},\ Dover Books on Mathematics\ (\bibinfo  {publisher} {Dover
  Publications},\ \bibinfo {year} {2013})\BibitemShut {NoStop}%
\bibitem [{\citenamefont {Conforti}\ \emph {et~al.}(2014)\citenamefont
  {Conforti}, \citenamefont {Cornu{\'e}jols},\ and\ \citenamefont
  {Zambelli}}]{conforti2014integer}%
  \BibitemOpen
  \bibfield  {author} {\bibinfo {author} {\bibfnamefont {M.}~\bibnamefont
  {Conforti}}, \bibinfo {author} {\bibfnamefont {G.}~\bibnamefont
  {Cornu{\'e}jols}}, \ and\ \bibinfo {author} {\bibfnamefont {G.}~\bibnamefont
  {Zambelli}},\ }\href {https://books.google.com/books?id=antqBQAAQBAJ} {\emph
  {\bibinfo {title} {Integer Programming}}},\ Graduate Texts in Mathematics\
  (\bibinfo  {publisher} {Springer International Publishing},\ \bibinfo {year}
  {2014})\BibitemShut {NoStop}%
\bibitem [{\citenamefont {Rozada}\ \emph {et~al.}(2019)\citenamefont {Rozada},
  \citenamefont {Aramon}, \citenamefont {Machta},\ and\ \citenamefont
  {Katzgraber}}]{PhysRevE.100.043311}%
  \BibitemOpen
  \bibfield  {author} {\bibinfo {author} {\bibfnamefont {I.}~\bibnamefont
  {Rozada}}, \bibinfo {author} {\bibfnamefont {M.}~\bibnamefont {Aramon}},
  \bibinfo {author} {\bibfnamefont {J.}~\bibnamefont {Machta}}, \ and\ \bibinfo
  {author} {\bibfnamefont {H.~G.}\ \bibnamefont {Katzgraber}},\ }\href
  {\doibase 10.1103/PhysRevE.100.043311} {\bibfield  {journal} {\bibinfo
  {journal} {Phys. Rev. E}\ }\textbf {\bibinfo {volume} {100}},\ \bibinfo
  {pages} {043311} (\bibinfo {year} {2019})}\BibitemShut {NoStop}%
\bibitem [{\citenamefont {Wang}\ \emph
  {et~al.}(2015{\natexlab{a}})\citenamefont {Wang}, \citenamefont {Machta},\
  and\ \citenamefont {Katzgraber}}]{PhysRevE.92.063307}%
  \BibitemOpen
  \bibfield  {author} {\bibinfo {author} {\bibfnamefont {W.}~\bibnamefont
  {Wang}}, \bibinfo {author} {\bibfnamefont {J.}~\bibnamefont {Machta}}, \ and\
  \bibinfo {author} {\bibfnamefont {H.~G.}\ \bibnamefont {Katzgraber}},\ }\href
  {\doibase 10.1103/PhysRevE.92.063307} {\bibfield  {journal} {\bibinfo
  {journal} {Phys. Rev. E}\ }\textbf {\bibinfo {volume} {92}},\ \bibinfo
  {pages} {063307} (\bibinfo {year} {2015}{\natexlab{a}})}\BibitemShut
  {NoStop}%
\bibitem [{\citenamefont {Wang}\ \emph
  {et~al.}(2015{\natexlab{b}})\citenamefont {Wang}, \citenamefont {Machta},\
  and\ \citenamefont {Katzgraber}}]{PhysRevE.92.013303}%
  \BibitemOpen
  \bibfield  {author} {\bibinfo {author} {\bibfnamefont {W.}~\bibnamefont
  {Wang}}, \bibinfo {author} {\bibfnamefont {J.}~\bibnamefont {Machta}}, \ and\
  \bibinfo {author} {\bibfnamefont {H.~G.}\ \bibnamefont {Katzgraber}},\ }\href
  {\doibase 10.1103/PhysRevE.92.013303} {\bibfield  {journal} {\bibinfo
  {journal} {Phys. Rev. E}\ }\textbf {\bibinfo {volume} {92}},\ \bibinfo
  {pages} {013303} (\bibinfo {year} {2015}{\natexlab{b}})}\BibitemShut
  {NoStop}%
\bibitem [{\citenamefont {Durstenfeld}(1964)}]{Durstenfeld1964CA}%
  \BibitemOpen
  \bibfield  {author} {\bibinfo {author} {\bibfnamefont {R.}~\bibnamefont
  {Durstenfeld}},\ }\href@noop {} {\bibfield  {journal} {\bibinfo  {journal}
  {Commun. ACM}\ }\textbf {\bibinfo {volume} {7}},\ \bibinfo {pages} {420}
  (\bibinfo {year} {1964})}\BibitemShut {NoStop}%
\bibitem [{\citenamefont {Fisher}\ and\ \citenamefont
  {Yates}(1948)}]{FisherYates}%
  \BibitemOpen
  \bibfield  {author} {\bibinfo {author} {\bibfnamefont {R.~A.}\ \bibnamefont
  {Fisher}}\ and\ \bibinfo {author} {\bibfnamefont {F.}~\bibnamefont {Yates}},\
  }\href@noop {} {{\selectlanguage {English}\emph {\bibinfo {title}
  {Statistical tables for biological, agricultural and medical research}}}}\
  (\bibinfo  {publisher} {Oliver and Boyd},\ \bibinfo {address} {London},\
  \bibinfo {year} {1948})\BibitemShut {NoStop}%
\bibitem [{\citenamefont {Rosenkrantz}\ \emph {et~al.}(1977)\citenamefont
  {Rosenkrantz}, \citenamefont {Stearns},\ and\ \citenamefont
  {Lewis}}]{Rosenkrantz1977}%
  \BibitemOpen
  \bibfield  {author} {\bibinfo {author} {\bibfnamefont {D.~J.}\ \bibnamefont
  {Rosenkrantz}}, \bibinfo {author} {\bibfnamefont {R.~E.}\ \bibnamefont
  {Stearns}}, \ and\ \bibinfo {author} {\bibfnamefont {P.~M.}\ \bibnamefont
  {Lewis}, \bibfnamefont {II}},\ }\href {\doibase 10.1137/0206041} {\bibfield
  {journal} {\bibinfo  {journal} {SIAM Journal on Computing}\ }\textbf
  {\bibinfo {volume} {6}},\ \bibinfo {pages} {563} (\bibinfo {year} {1977})},\
  \Eprint {http://arxiv.org/abs/https://doi.org/10.1137/0206041}
  {https://doi.org/10.1137/0206041} \BibitemShut {NoStop}%
\end{thebibliography}%

\end{document}